 \documentclass[preprint2]{aastex}

\slugcomment{submitted to PASP}

\shorttitle{}
\shortauthors{}

\begin{document}


\title{Probing the Substellar Regime with SIRTF}


\author{Eduardo L. Mart\'\i n, Wolfgang Brandner, David Jewitt, Theodore Simon, and 
Richard Wainscoat}
\affil{Institute for Astronomy, University of Hawaii, 2680 Woodlawn 
Drive, Honolulu, HI 96822}
\email{ege@ifa.hawaii.edu,brandner@ifa.hawaii.edu,jewitt@ifa.hawaii.edu,simon@ifa.hawaii.edu,
rjw@ifa.hawaii.edu}

\author{Mark Marley}
\affil{Space Physics Research Institute, NASA Ames Research Center, Moffett Field, 
CA 94035}
\email{mmarley@mail.arc.nasa.gov}

\and

\author{Christopher Gelino}
\affil{Department of Astronomy, New Mexico State University, Las Cruces, NM 880033}
\email{crom@nmsu.edu}



\newpage 

\begin{abstract}
One of the main scientific drivers of the Space InfraRed Telescope 
Facility (SIRTF) is the
search for brown dwarfs and extrasolar superplanets.
We discuss observational strategies for identification of these objects, 
and conclude that an optimal strategy is a wide IRAC survey (18~deg$^2$) 
with a 5~$\sigma$ sensitivity of 3.9~$\mu$Jy in channel 2 
(M$\sim$19.1$^m$). For this sensitivity, we provide 
estimates of the 
number of low mass brown dwarfs and isolated planets detected per square degree 
for power-law mass functions with $\alpha$=1.5, and 1.0. 
Shallower surveys covering a larger area are inefficient because of large 
overheads and detector noise. Deeper surveys covering 
a smaller area become more and more affected  
by crowding with galaxies. 
A survey like the one that we propose would determine  
the field mass function down to a few
Jupiter masses through the identification of a large sample
of brown dwarfs and isolated planets.  
The proposed SIRTF survey would also allow the first 
detection of ultracool substellar objects with temperatures
between 700~K and 200~K. The cooling curves of substellar objects 
with masses less than 20 Jupiters imply that they should spend most 
of their lifetimes at temperatures below 700~K. Preliminary
models indicate that their atmospheres could be dominated by
water clouds, which would diminish their optical and near-infrared 
fluxes. The properties of those objects are still
completely unexplored.

\end{abstract}




\section{Introduction}

The past five years have witnessed the remarkable birth of a new 
research field of astronomy --- the physical study of substellar objects 
outside the solar system. 
Large scale optical 
and near-infrared surveys of the sky (DENIS, 2MASS and SDSS) have revealed very cool 
objects that are close to us but nevertheless were unknown.  
The overwhelming majority of substellar objects in the field, however, cannot be
detected by ground-based surveys; such objects 
cool very rapidly with time and become extremely faint at wavelengths shorter
than about 4~$\mu$m.  SIRTF (Fanson et al. 1998) will allow to 
probe deeper in the substellar regime, extending the census of 
substellar objects in the field 
to lower masses and temperatures (see Figure~1). 

The separation between brown dwarfs and planets is a matter of convention. 
We concur with other authors (e.g., Basri \& Marcy 1997; Oppenheimer et al. 2000) 
in arguing that a definition based on the well known laws of gravitation 
and nuclear physics is more reliable than a definition based on the origin 
of the objects. Thus, we adopt the following 
definitions: A brown dwarf is an object unable to stabilize on the 
hydrogen burning main-sequence, but able to fuse deuterium into helium. 
For solar metallicity the mass range of brown dwarfs is 
between 0.075~M$_\odot$ and 0.013~M$_\odot$ 
(Baraffe et al. 1998, Saumon et al. 1996). Planets are objects unable to fuse 
deuterium, i.e., they have masses lower than 0.013~M$_\odot$. 
``Isolated  planets'' are planets that are not bound to any particular star. 
Recent results suggest that this kind of planets may be numerous 
in the disk of the Galaxy (B\'ejar et al. 2000; Zapatero Osorio et al. 2000). 
The smooth transition of the mass function from the brown dwarfs to the isolated planets indicates 
that they have a common origin, which is much like that of star formation from fragmentation and collapse of 
molecular clouds, but not like the planets in our solar system (accretion in a protoplanetary disk).  
Despite of their different origins, isolated planets may have the same basic properties (overall chemical composition, 
luminosity, temperature) as their counterparts orbiting stars. 

Studies show that the number of substellar objects increases towards 
lower masses. The functional form of this relationship, the Initial 
Mass Function (IMF hereafter), can be approximated by a power law 
dN/dM $\sim$ M$^{-\alpha}$ (Salpeter 1955). For masses between 
0.1~M$_\odot$ and 0.01~M$_\odot$ (i.e., $\sim$ 100 and 10~M$_{Jup}$, respectively), 
several surveys in young open clusters have found $\alpha=1.0\pm0.5$.
(B\'ejar et al. 2000, Hillenbrand \& Carr 2000; Mart\'\i n et al. 2000; Najita et al. 2000).
The density of field brown dwarfs discovered by DENIS and 2MASS is also 
consistent with this mass function (Reid et al. 1999). 
The number of substellar objects in each cluster, or in the field, is still small. 
Hence the statistical uncertainties are large and it is not possible 
to conclude if there are significant variations in the IMF from one 
region to another. The cooling curves of brown dwarfs and planets imply that 
they spend a large part of their lives at extremely low luminosities and cool temperatures. 
The launch of the SIRTF satellite provides an unprecedented opportunity for detecting 
ultracool substellar objects. 

This paper is organized as follows: Section 2 gives a brief description of 
the expected spectra of ultracool substellar objects and justifies 
the need for SIRTF to detect them; Section 3 discusses observational strategies 
to detect substellar objects with SIRTF; Section 4 
provides estimates of the counts of substellar objects in the disk that can be 
detected with SIRTF; and Section 5 summarizes the main conclusions.

\section{The need for SIRTF: Infrared Spectra of Ultracool Substellar Objects}

We have used calculations for solar composition substellar objects. 
These models will be described in detail in a forthcoming paper 
(Marley et al. 2001, in preparation). We limit this presentation to a brief 
discussion of results that are relevant for our purposes. 

With decreasing 
temperature the molecular bands of methane and water depress the flux 
at optical and near-infrared wavelengths. At temperatures below $\sim$700~K, 
substellar objects emit very little flux shortward of 4~$\mu$m (see 
Figure~2). In fact, the coolest known brown dwarfs, namely  
Gl~570~D, has a temperature of 790$\pm$50~K (Burgasser 
et al. 2000). Its spectrum is similar to that of Gl~229~B. 

It is unlikely that the DENIS and 2MASS surveys will detect 
any objects significantly cooler than Gl~570~D.  With J$\ =15\fm33$ and M$_J\ 
=16\fm5$, this object is near the detection limit of 2MASS (J\,$\sim16^m$; 
and, J\,$\sim15^m$ for DENIS).
A substellar object 200~K cooler would have an absolute J-band magnitude
about 1.5 mag fainter (Burrows et al. 1997), and to be detected by these
surveys it would have to be 
2 times closer, i.e., at 3~pc. There are 6 stellar systems within  
3~pc of the Sun, and there could be a similar number of brown dwarfs. 
However, the chances that 2MASS will detect them may be negligible if the 
near-infrared spectrum is depleted by strong water bands. 

Preliminary models (Marley 2000), illustrated here 
in Figure~3, provide a glimpse of what these substellar objects may look 
like. Water clouds will first appear in objects with $T_{\rm eff}$ below 
$\sim$600~K. The clouds will 
first form as tenuous stratus clouds high in the atmosphere and then 
form lower in the atmosphere in cooler objects where they will be
thicker and more substantial.  For $T_{\rm eff}$ below $\sim$300~K, 
ammonia clouds will form. Jupiter's water clouds are hidden below two 
higher cloud layers and are poorly understood.  Thus SIRTF provides 
the opportunity to detect the first substellar mass objects outside
our solar system in which water clouds dominate the atmosphere.

Water clouds will substantially alter the infrared spectra
of these objects.  T-dwarfs like Gliese 229 B are blue
in J-K.  Once water clouds form, the J-K color will move
redward towards J-K ~ 0 to 1 (Marley and Ackerman 2000). 
A new spectral class, P for example, could be necessary for classifying 
these objects cooler than the T dwarfs. 
Based on near-IR colors (JHK) alone, it will be very difficult
to distinguish the P-type brown dwarfs from background M-stars
with similar colors.
Brown dwarfs with clear atmospheres, like the T-dwarfs, are bright
at 4 to 5 um (Marley et al. 1996; Burrows
et al. 1997).  As with Jupiter, Marley and Ackerman (2000)
find that the coolest brown dwarfs are also bright
in this spectral region, even when water clouds are present (Figure~3).
Thus a search with SIRTF at 4-5um has the potential of
easily discovering such objects. 
 
As illustrated in Figure~2, the IRAC (Fazio et al. 1998) passband at 4.5~$\mu$m (channel 2) 
offers the best sensitivity for detection of a 10 M$_{Jup}$\ object, particularly for old ages.
2MASS can detect a 10 M$_{Jup}$\ isolated planet only if it is about 0.1~Gyr old 
and is located at a distance of less than 10 pc. It is unlikely that such an object 
exists.  On the other hand, SIRTF can provide a 5~$\sigma$ detection in 720~s of 
a 10 M$_{Jup}$\ isolated planet 
with an age of 0.1~Gyr out to 350~pc at 4.5~$\mu$m and 200~pc at 3.6~$\mu$m. 
SIRTF can also detect a 10 M$_{Jup}$\ isolated planet at an age 1~Gyr up to a distance of
120~pc at 4.5~$\mu$m and 40~pc at 3.6~$\mu$m. 

\section{Observational Strategies for Detecting Substellar Objects with SIRTF} 

 The DENIS, 2MASS and SDSS surveys have demonstrated that free-floating brown dwarfs 
outnumber brown dwarf companions to stars by about a factor of 10.
These surveys combined have already discovered more than 200 free-floating
brown dwarfs (Burgasser et al. 1999; Delfosse et al. 1999; Fan et al. 
2000; Gizis et al. 2000; Kirkpatrick et al. 2000; Leggett et al. 2000; 
Mart\'\i n et al. 1999),
but only three companion brown dwarfs to stars (Burgasser et al. 2000;
Kirkpatrick et al. 2000). Moreover, high-precision radial velocity surveys have 
revealed a brown dwarf desert within a few AU of G and K-type main-sequence 
stars  (Halbwachs et al. 2000). They determine that there exists
a secondary mass cutoff between 0.01 and 0.1 M$_\odot$. Other searches for
brown dwarf companions to stars have found only a few outstanding examples 
in samples of hundreds of stars (Becklin \& Zuckerman 1988; 
Oppenheimer 1999). These companions are very interesting but seem to be rare. 

 The spatial resolution of SIRTF (FWHM=2".3 with IRAC), and the lack of a 
coronograph, will not allow a Jupiter mass planet at 5 AU separation to be 
resolved, even for the closest stars. Jupiter-mass planets may be common 
around stars, but they are unlikely to be found outside a 20~AU radius of 
the parent star. The ice condensation region is located at about 5~AU, with 
little dependence on the spectral type of the star, and gravitational 
instabilities in the disk are unlikely to be effective beyond 20 AU
(Boss 2000). 
Direct images of massive disks around T Tauri stars 
indicate that in general the density is too low to allow for the
formation of a giant planet beyond 20 A.U.
(e.g., Krist et al. 2000). On the other hand, the
deep surveys of Orion and IC~348 suggest that free-floating planets 
probably are very numerous (Najita et al 2000; Zapatero Osorio et al. 2000).  

These considerations lead us to propose that a wide field imaging survey with 
SIRTF is the best strategy to detect a large sample of substellar objects. 
We have used the SIRTF planning observations tool (SPOT) to simulate future observations. 
With total exposure times of 720~s per pointing (divided into 24 exposures of 30~s for 
dithering and cosmic ray removal), 
we could make an IRAC map covering 1 square degree in 33.3 hours of observation time. 
The 5~$\sigma$ sensitivity for point sources under conditions of low background are:   
2.91~$\mu$Jy in channel 1, 3.88~$\mu$Jy in channel 2, 11.3~$\mu$Jy in channel 
3, and 13.2~$\mu$Jy in channel 4.
At these flux limits, the main contribution to source
confusion comes from background galaxies
We have estimated the number of galaxies
and stars at the IRAC passbands using the predictions of Franceschini et al. (1991) and 
a modified version of the galaxy model of
Wainscoat et al. (1992).  
 the recent 7$\mu$m ISOCAM counts obtained by Taniguchi et al. (1997). 
Our calculations show that in IRAC channel 2 (our most sensitive passband for substellar objects) 
we expect to detect approximately 3,800 stars (for high galactic latitude) and 
78,000 galaxies per square degree. 
If each galaxy is assumed to occupy
$3\times3$ pixels ($3.6\times3.6$ arcsec), then the galaxies will cover approximately 7\% of
the pixels.  At fainter
detection limits, the crowding from galaxies is
correspondingly greater.  Thus, more sensitive
observations offer relatively diminishing returns 
for a much greater investment of observing time.

%\section{Identifying Substellar Objects with SIRTF}

According to the
fluxes predicted by our synthetic spectra, 
substellar objects can be identified unambiguously
from IRAC photometry. 
In fact, it is enough with a 5~$\sigma$ detection in channel 2 and 
no detection in the other channels. As shown by Figure 4, any object
with log F(4.5$\mu$m)/F(3.6$\mu$m) $>$ 0.3 can be considered as a good
brown dwarf or planet candidate.
Neither stars nor galaxies contaminate the substellar locus. 
The location of flared and flat disks in Figure~4 is based on results on the spectral
index for T Tauri star disks by Kenyon \& Hartmann (1995). The
location of 30 Doradus has been derived from ISO SWS measurements
by Sturm et al. (2000), and the flux values for L$^*$ galaxies at z=0.1 to 5
are based on simulations by Fazio et al. (1998).

Figure~4 also indicates that
Population III\footnote{Population III stars in general refers to the first generation of
stars, i.e. objects which formed at zero-metallicity (e.g. Cojazzi
et al. 2000, and references therein). Similarly, Population III
brown dwarfs refers to substellar objects, which formed at zero-metallicity.} 
brown dwarfs with effective temperatures below 400\,K are 
readily identified from their colors alone. Still warmer (and more massive) 
Population III brown dwarfs occupy a region in the color-color plot that 
coincides with redshifted L* galaxies and with stars having debris disks 
which are intermediate between Vega-type disks and the disks around classical 
and weak-line T\,Tauri stars. Additional information will therefore be needed 
to make a positive identification. Two epochs of IRAC observations would  
facilitate an identification of the most nearby Population III brown dwarfs,
based on their high proper motion (and possibly their parallax). More distant
candidates for massive Population III brown dwarfs will have to be followed 
up with instruments with higher instrinsic angular resolution, e.g., 
ground-based 8-m to 10-m class telescopes, the Next Generation Space Telescope 
(NGST), the Terrestrial Planet Finder (TPF), or the Stratospheric Observatory 
for Infrared Astronomy (SOFIA).

\section{Modelling the Number Counts of Substellar Objects in the Disk}

We have estimated the number counts of substellar objects in the disk of the Galaxy  
using the following ingredients: 
(1) Theoretical fluxes computed from synthetic spectra (see Figure 2). 
(2) Sensitivities for low-background from the IRAC manual. We have adopted a 5$\sigma$ confidence 
level for our  detection limit in each channel.     
(3) A solar neighborhood density of 0.057 stellar systems/pc$^3$ (Reid et al. 1999). 
(4) Power-law mass functions with $\alpha$=1.5, 1.0 and 0.5 from 100 to 1 M$_{Jup}$\ . 
(5) A constant star formation rate over the lifetime of the Galactic disk (from 0 to 10~Gyr). 
(6) An exponential vertical distribution for distances beyond 50~pc, and a scale height of 325~pc, 
consistent with the counts of M dwarfs.  

A Monte Carlo code provided the cumulative number of substellar objects. 
In Figure~5, we show the cumulative 
distributions of objects with masses in the range 25 to 1 M$_{Jup}$ for two different 
slopes of the IMF power-law. 
For an IMF with $\alpha$ = 1.5, we predict that SIRTF could detect about 30 isolated planets 
per square degree. For $\alpha$ = 1.0, the number of planets decreases to about 4 per square degree.

We consider the hypothetical case that 600 hours of SIRTF time would 
be dedicated for an IRAC survey to search for substellar objects. 
Using SPOT we have estimated that an area of 18 square degrees could be covered with 
5~$\sigma$ sensitivity of 3.88~$\mu$Jy in channel 2. 
The cumulative distribution of substellar objects that such a survey 
would find (for $\alpha$ = 1.5) is given in Figure~6. We compare with 
another survey with which is less sensitive but covers a larger area (64~deg$^2$). 
Both surveys require the same amount of SIRTF observing time according to the SPOT program. 
The larger survey is less efficient due to larger overheads and the 
contribution of detector noise when the exposure times are shorter. 
More than 150 substellar 
objects with temperatures cooler than 600~K could be revealed in 
18 square degrees. If the 
IMF is shallower the number of objects would naturally be less. 
For $\alpha$ = 1.0 we predict about 20 objects cooler than 600~K, 
and for $\alpha$ = 0.5 we predict only three of them.  
SIRTF observations can constrain the field mass function down to masses below the deuterium limit 
by counting the number of ultracool substellar objects.

\section{Conclusions} 

We have shown that IRAC channel 2 is very efficient for detecting ultracool substellar objects. 
We present an optimized observational strategy for detection of brown dwarfs and isolated planets 
with temperatures below 700~K. Preliminary models indicate that the atmospheres of these objects 
are dominated by water clouds, which make the objects fainter at optical and near-infrared wavelengths 
than previously thought. The 5~$\mu$m window is an excellent region for detecting them because 
their atmospheres are more transparent. 
Our strategy consists of a SIRTF survey with a 5~$\sigma$ sensitivity 
of 3.88~$\mu$Jy in IRAC channel 2. We show that substellar objects can easily be identified in 
an IRAC color-color diagram. Any object
with log F(4.5$\mu$m)/F(3.6$\mu$m) $>$ 0.3 can be considered as a good
brown dwarf or planet candidate.
Neither stars nor galaxies contaminate the substellar locus. 
With our detection limit, we predict a density of 
30 isolated planets per square degree for an IMF with a slope $\alpha$ = 1.5. In 600 hours of 
SIRTF time, an area of 18 square degrees would be covered. We predict that such a survey would find 
about 540 isolated planets, and about 160 substellar objects with temperatures cooler than 600~K. 
Such objects cannot be detected with the current generation of near-infrared all-sky surveys 
(DENIS and 2MASS).  For shallower IMF slopes, the number of substellar objects diminishes. Thus, the 
proposed survey can determine the field substellar IMF down to very low masses.

\acknowledgments

We would like to thank the following people for useful comments and discussions:  
Xavier Delfosse, Thierry Forveille, John Gizis, Pat Henry, Klaus Hodapp, Joseph Hora, 
Jeff Kuhn, Alan Tokunaga, Motohide Tamura, and Maria Rosa Zapatero Osorio. 
We thank France Allard and Isabelle Baraffe for sending their theoretical models so that 
we could compare them with the models used in this paper.




%% Generally speaking, only the figure captions, and not the figures
%% themselves, are included in electronic manuscript submissions.
%% Use \figcaption to format your figure captions. They should begin on a
%% new page.

\clearpage

%% No more than seven \figcaption commands are allowed per page,
%% so if you have more than seven captions, insert a \clearpage

%% There must be a \figcaption command for each legend. Key the text of the
%% legend and the optional \label in curly braces. If you wish, you may
%% include the name of the corresponding figure file in square brackets.
%% The label is for identification purposes only. It will not insert the
%% figures themselves into the document.
%% If you want to include your art in the paper, use \plotone.
%% Refer to the on-line documentation for details.

\figcaption[fig1.eps]{Parameter space occupied by Brown Dwarfs and Planets. Objects
were selected from an $\alpha$ = 1.0 power law distribution in mass
with 0.01 $<$ age $<$ 10 Gyr.  Ground-based surveys are sensitive only
to the minority of objects with T$_{\rm eff} >$ 700~K.  SIRTF, by contrast,
is sensitive to the far more numerous objects at lower temperatures. \label{fig1}}

\figcaption[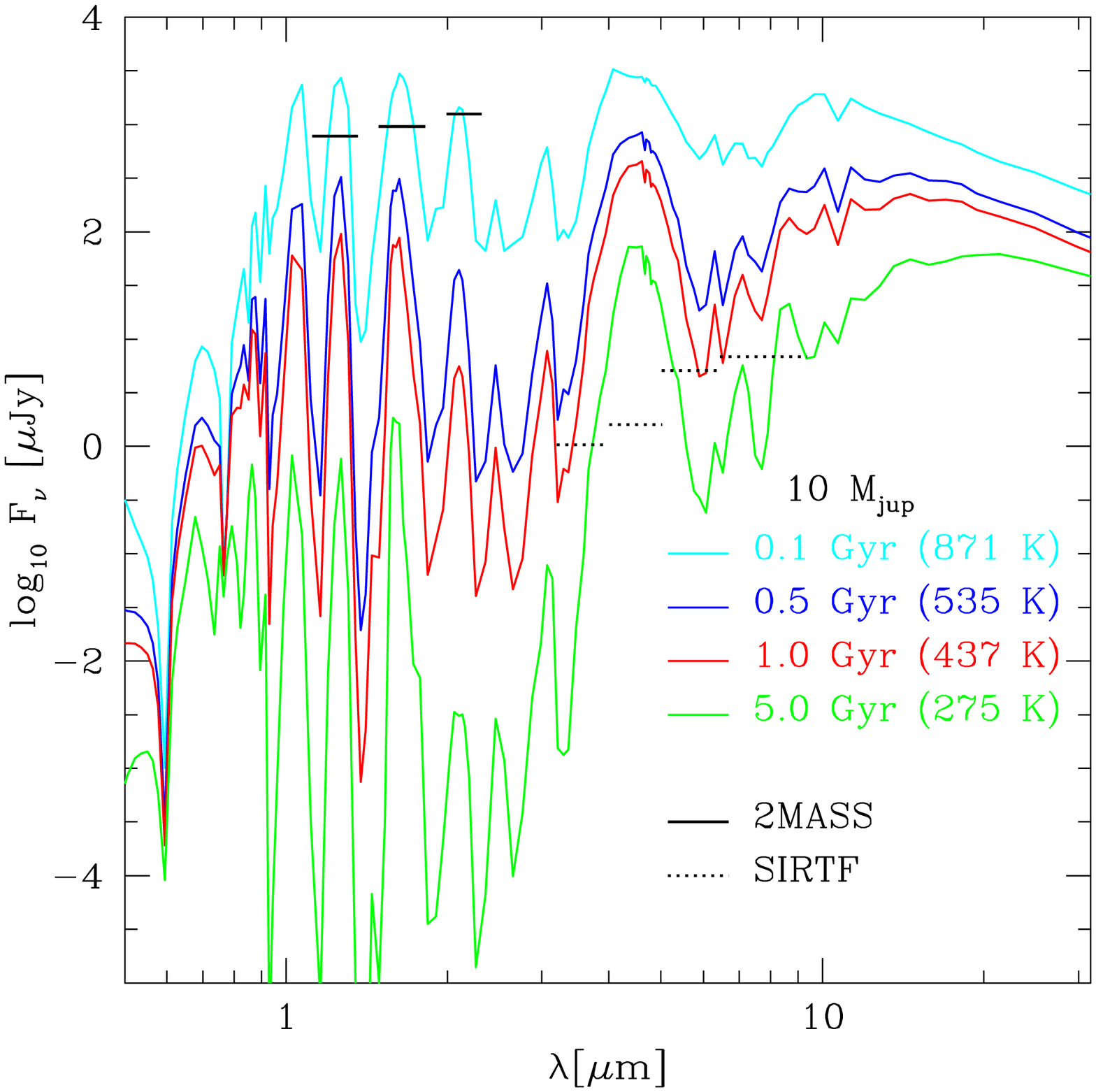]
{Synthetic spectra of isolated superplanets with 10 Jupiter mass 
at 10~pc from the Sun and a variety of ages. The detection limits of 2MASS (J, H and 
K$_s$) and our proosed SIRTF/IRAC survey are overplotted.  \label{fig2}}

\figcaption[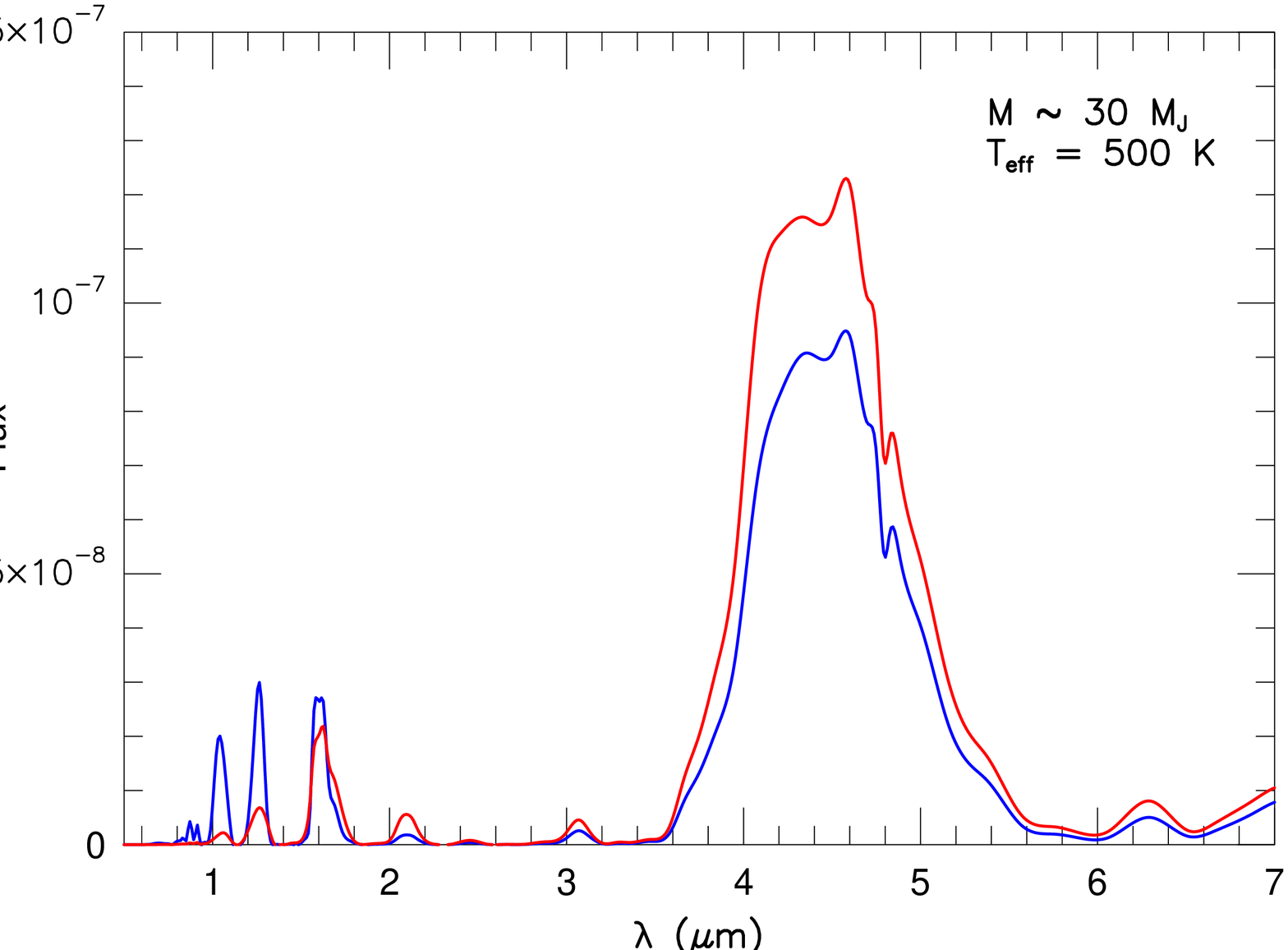]
{Smoothed model spectra for clear (blue) and cloudy (red) 30
M$_{\rm Jup}$ brown dwarf on arbitrary, linear flux scale (Marley 2000).
Note the prominence of the $4-5\mu$m flux window.  The clear model, like
Gliese 229 B, is blue in J--K ($J-K\sim-1.6^m$) whereas the cloudy model
is red ($J-K\sim 0.6^m$).  Details of spectra are dependent upon
uncertain cloud properties and will provide substantial information 
on cloud physics in a substellar mass atmosphere (Ackerman \& Marley 
2000). Flux units are in erg cm$^{-2}$ sec$^{-1}$ Hz$^{-1}$. \label{fig3}} 

\figcaption[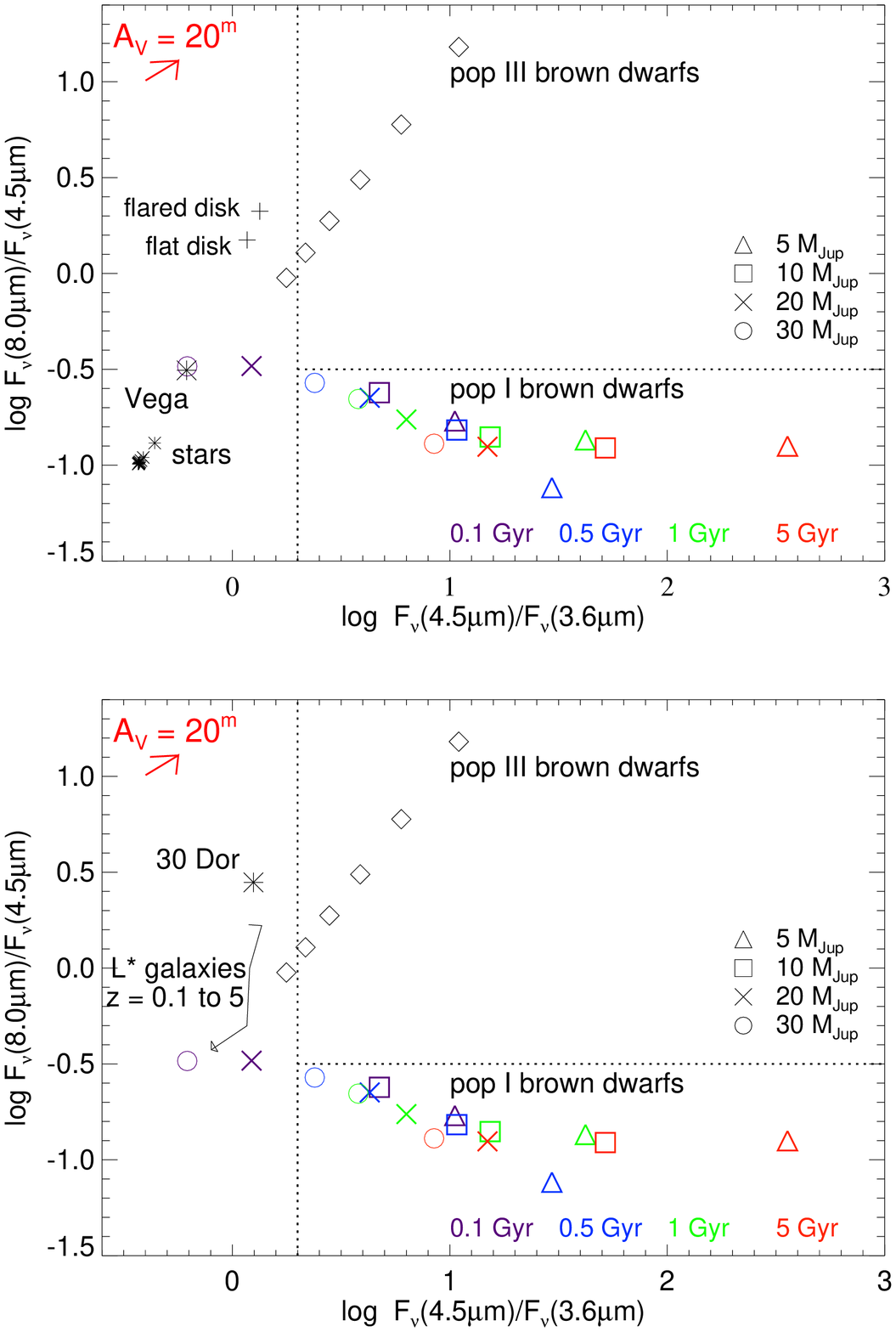]
{In color-color space, brown dwarfs are clearly distinct from stellar
sources (top panel) and extragalactic sources (lower panel).\label{fig4}}

\figcaption[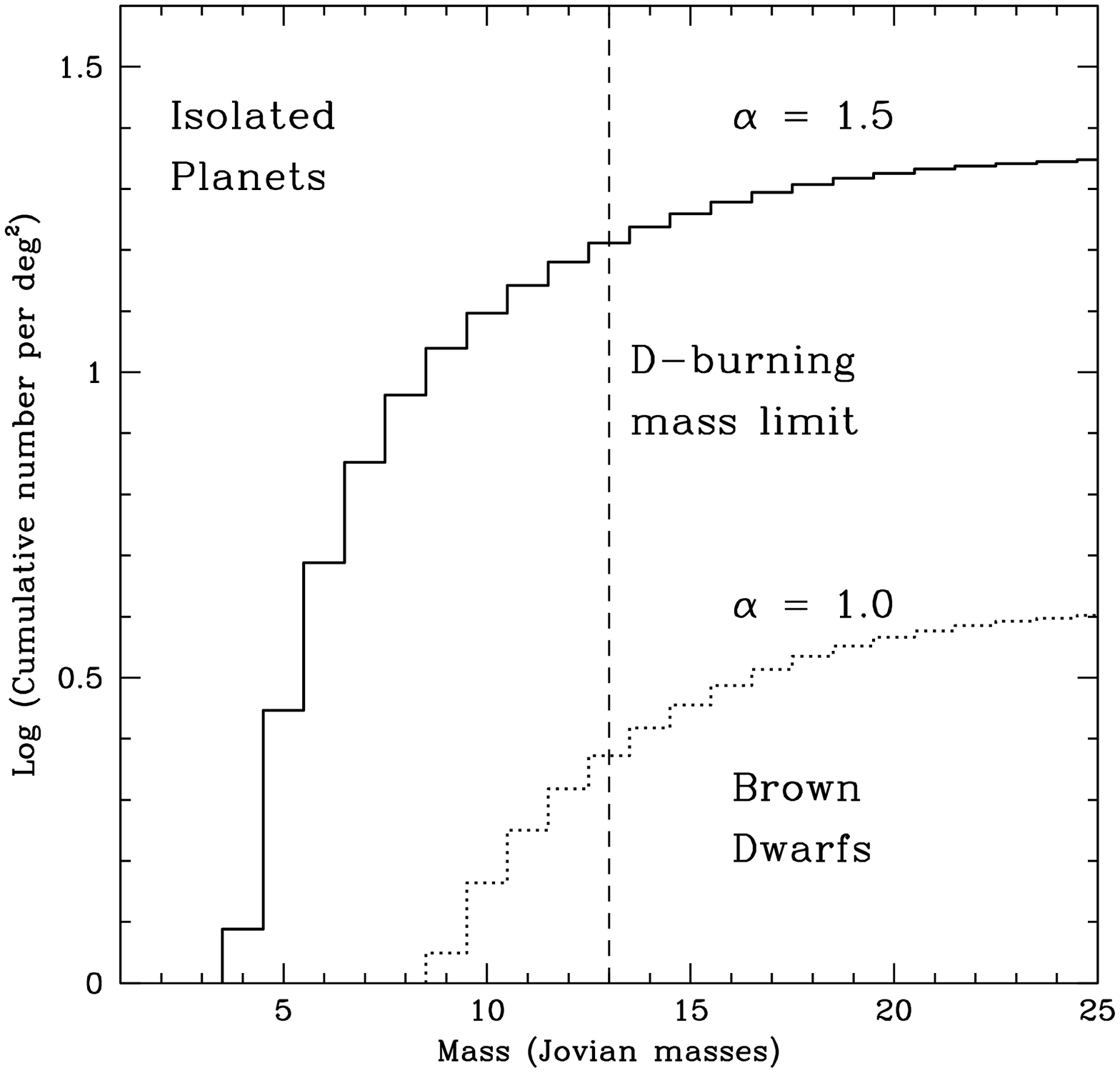]
{Cumulative number of low-mass brown dwarfs and isolated 
superplanets that could be identified in our 
survey. Solid histrogram corresponds to a power-law IMF with $\alpha$ = 1.5; Dotted 
histrogram corresponds to a power-law IMF with $\alpha$ = 1.0. \label{fig5}}

\figcaption[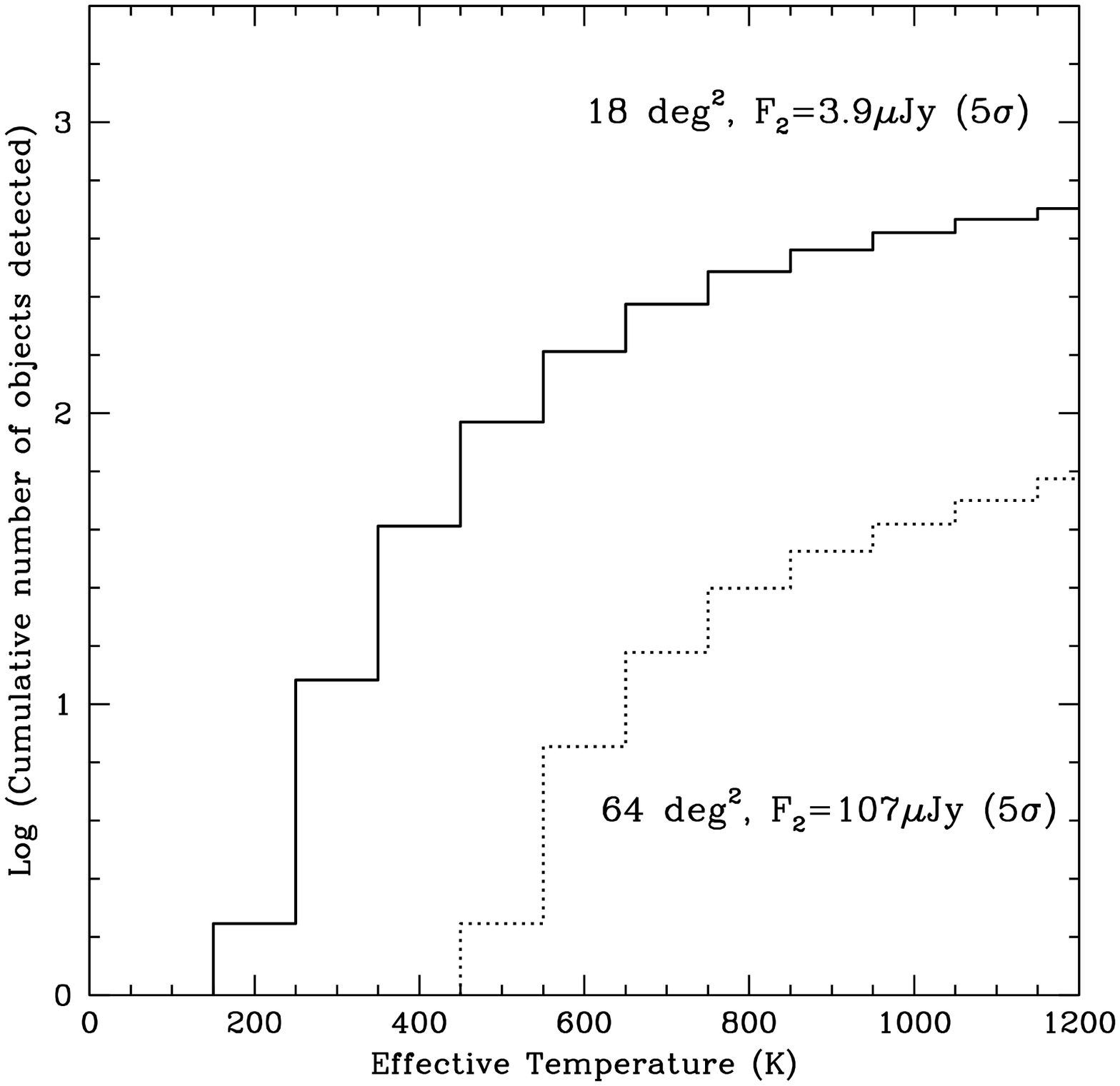]
{Cumulative number of brown dwarfs and isolated superplanets identified at  
5$\sigma$ limiting flux in channel 2 as a function of effective temperature.  We compare with 
a hypothetical survey that would require the same amount of SIRTF time 
(600 hours), covering more area but with less sensitivity. We have used an IMF with 
$\alpha$=1.5.\label{fig6}}

\end{document}